\documentclass[twocolumn,showpacs,preprintnumbers,amsmath,amssymb]{revtex4-1}
\usepackage{dcolumn}
\usepackage{color}
\usepackage[normalem]{ulem}
\usepackage{amssymb,amsmath,amsthm,epsfig,subfigure} 
\usepackage{bm}
\def\beq{\begin{equation}}
\def\eeq{\end{equation}}
\def\beqar{\begin{eqnarray}}
\def\eeqar{\end{eqnarray}}

\newcommand{\N}{|N|}
\newcommand{\ho}{\hat{\omega}}
\newcommand{\hphi}{\hat{\varphi}}

\newcommand{\U}{{\bf U}}

\newcommand{\curl}{{\nabla \times}}

\newcommand{\x}{{\bf x}}

\newcommand{\kk}{{\bf k}}

\newcommand{\tn}{{\tilde {\rho}}}

\newcommand{\p}{\partial}

\newcommand{\vphi}{{\tilde {\varphi}}}
\newcommand{\tdomega}{{\widetilde {\delta \omega}}}

\newcommand{\gapprox}{\lower.4ex\hbox{$\;\buildrel
>\over{\scriptstyle\sim}\;$}}
\newcommand{\lapprox}{\lower.4ex\hbox{$\;\buildrel
<\over{\scriptstyle\sim}\;$}}

\begin{document}
\title{Effects of shear flows on the evolution of fluctuations in interchange turbulence}

\author{Ismail Movahedi and Eun-jin Kim}
\affiliation{
School of Mathematics and Statistics, University of Sheffield, Sheffield, S3 7RH,
U.K.}

\begin{abstract}
We report a non-perturbative study of the effect of different type of shear flows on the evolution of vorticity and particle density fluctuations in interchange turbulence. For the same shear strength, the transport of density is less reduced by streamers than by zonal flows, zonal flows leading to oscillation death. In the inviscid limit, vorticity (density) grows (decays) as a power law due to streamer or zonal flow, and exponentially due to the combined effect of zonal flow and streamer with the same sign of shear. Zonal flow and streamer with the opposite sign of shear lead to oscillation at multiple frequencies. 

\end{abstract}. 

\maketitle


Shear flows play a primary role in turbulence regulation and transport quenching in a variety of systems \cite{BURRELL,PLANET,ATM,HUNT,KN06,EXP,GARBET,DAM2017,Chang2017,KIM1}. This effect is particularly important in magnetically
confined fusion plasmas (see, e.g., \cite{BURRELL,EXP,GARBET,DAM2017,Chang2017,KIM1}) as shear flows tend to be stable.
Since the discovery of the low-to-high (L-H) transition \cite{ASDEX}, a great effort has been
devoted to elucidating the detailed dynamics involved in the formation of transport barrier by different shear flows
in different models.
 
In the case of coherent shearing by mean ${\bf E} \times {\bf B}$ flows, the transport of passive scalar fields
was shown to be reduced as ${\Omega}^{-1}$, where ${{\Omega}}$ is the mean shearing rate \cite{KIM1}. However,
a significant reduction in particle (heat) transport was found in simple interchange (ion-temperature gradient)
turbulence model \cite{KDH}. This result was based on the framework where the feedback of the velocity 
on density fluctuation was not included but instead was treated as a source of free energy and thus as a part of a stochastic
noise that drives the density fluctuation. The coupling between density and velocity fluctuations through this feedback
however causes not only the excitation of waves (e.g. gravity waves) or instability, depending on the relative direction of
the background density gradient and the effective gravity, but also very peculiar dynamics, as we will show.

In this Brief Communications (BC), we report the effects of shear flows on the evolution of the vorticity and particle density 
fluctuations in interchange turbulence by treating the coupling between the vorticity and particle density fluctuation consistently.
To elucidate the effects of different types of shear flows, we consider i) zonal flows, ii) streamers and iii) combined zonal flows and streamers. Streamers \cite{CHAMP} are radially elongated convective cells (i.e., polodially localized, radial flows) and thus can directly contribute to radial transport by advection. Also, unlike zonal flows, streamers can be excited by a linear instability. 
On the other hand, streamers can lead to turbulence quenching due to shear, like zonal flows. 
While the generation of streamers has been studied in different (drift, interchange, RT, etc) models \cite{CHAMP,Das}, much less work was done on the effect of streamer shear on turbulence suppression and time-variability. For instance, the effect of zonal flow and streamer shears on turbulence is simply assumed  to be similar \cite{Zhu}. To highlight the different effects of shear flows, we focus on the initial value problem.
 
\noindent
{\bf Model}: In interchange turbulence model with cold ions (which is similar to the classical Rayleigh-B\'enard convection
problem (e.g. \cite{KN06})), we consider the quasi-linear evolution of flute-like perturbations of particle density $\rho$ and vorticity 
$\delta \omega = \curl {\bf v}$ in the two dimensional (2D) $x-y$ plane \cite{KDH}. Here, $x$ and $y$ represent the local radial 
and poloidal directions, respectively, perpendicular to a magnetic field  ${\bf B} = B_0 {\hat z}$; 
$\langle {\bf v} \rangle = \langle \delta \omega \rangle = \langle \rho \rangle = 0$, where the angular brackets denote
the average. By taking the effective gravity in the radial direction as ${\bf g}=g {\hat x}$ (due to magnetic curvature, etc), we have
\begin{eqnarray}
{\p_t} \delta \omega + \U \cdot \nabla \delta \omega  = -g \p_y{\rho}/\rho_m + 
\nu \nabla^2 \delta \omega \,,
\label{eq2}\\
{\p_t} \rho + \U \cdot \nabla \rho  = -v_x \p_x \rho_0 + D \nabla^2 \rho \,.
\label{eq1}
\end{eqnarray}
Here, $\rho_{m}$ is the constant background density, $\rho_0(x)$ is the mean background density and $\rho$ is the fluctuation.
${\bf u } = {\bf v}+\U$ is the total velocity, consisting of fluctuation ${\bf v} = -(c/B_0)\nabla \phi \times {\hat z}$
and the mean flow $\U$. We take $\U$ to be either zonal flow only $\U=(0, -x A_z)$, streamers only $\U = (-y A_s, 0)$, 
or combined zonal flows and streamers $\U = (-y A_s, -x A_z)$.
$D$ and $\nu$ capture the coherent nonlinear interaction (i.e., ``eddy diffusivity and viscosity'') as well as molecular
dissipation. Unlike the previous work \cite{KDH} where the source of free energy $v_x \p_x \rho_0$ in Eq.\ (\ref{eq1}) was treated as a part of the noise, we investigate the consequence of the coupling between $\delta \omega$ and $\rho$ through the buoyancy
by solving Eqs. (\ref{eq2}) and (\ref{eq1}) simultaneously.
We recall that this coupling term determines the stability, giving rise to stable or unstable gravity waves
depending on whether $\frac{\partial \rho_0}{\partial x} > 0$ (density increasing in the direction of gravity) 
or $\frac{\partial \rho_0}{\partial x} <0$, respectively. In this BC, we focus on the stable case with $\nu=D$; 
similar results would follow in the unstable case.

In order to capture the effect of distortion of an eddy (i.e., wind-up)  by $\U$ non-perturbatively, we employ the time-dependent wavenumber $\kk$:
\begin{equation}
\rho({\bf x},t) = \tilde{\rho}({\bf k},t) \exp{\{i(k_x(t) x + k_y(t) y)\}}\,,
\label{eq3}
\end{equation}
and similarly for ${\bf v}$ and $\delta \omega$.
Plugging Eq. (\ref{eq3}) in Eq. (\ref{eq2}) and (\ref{eq1}), we find that we can eliminate the advection term by $\U$ if
$(\p_t \kk) \cdot \x + \U \cdot  \kk = 0$, which is
\begin{eqnarray}
\p_t k_x(t) =  A_z k_y,\,\,\,\,\, \p_t  k_y(t)= A_s k_x
\label{eq6}
\end{eqnarray}
for $\U = (-y A_s, -x A_z)$.
By using Eqs. (\ref{eq3}) and (\ref{eq6}),  we recast Eqs. (\ref{eq2})-(\ref{eq1}) as follows:
\begin{eqnarray}
\partial_{t} \ho & = &k_y \hphi,
\label{eq9}\\
\partial_{t } {\hphi} &= &- \frac{k_y N^2}{k_x^2 + k_y^2} \ho.
\label{eq10}
\end{eqnarray}
Here,  $\vphi= -\frac{ig}{\rho_m} \tn$, $\ho(t)  = e^{\nu Q(t)} \tdomega$, $\hphi =  e^{\nu Q(t)} \vphi$
and $Q(t) = \int_0^t dt_1\,[k_x(t_1)^2 + k_y(t_1)^2]$;
$N^2 = \frac{g}{\rho_{m}} \frac{\partial \rho_{0} }{\partial x}$ is the square of the buoyancy frequency.\\

%

\noindent
{\bf Zonal flow only}: To understand the effect of the coupling between $\tn$ and $\tdomega$, we start with the case of zonal flow only $\U =(0,  -xA_z)$,
in which case Eq. (\ref{eq6}) becomes
\begin{equation}
k_x(t) = k_x(0) + k_y(0)A_zt,\,\,\,\,
k_y(t) = k_y(0).
\label{eq11}
\end{equation}
\begin{figure}
	\centering
	\includegraphics[height=2.5in]{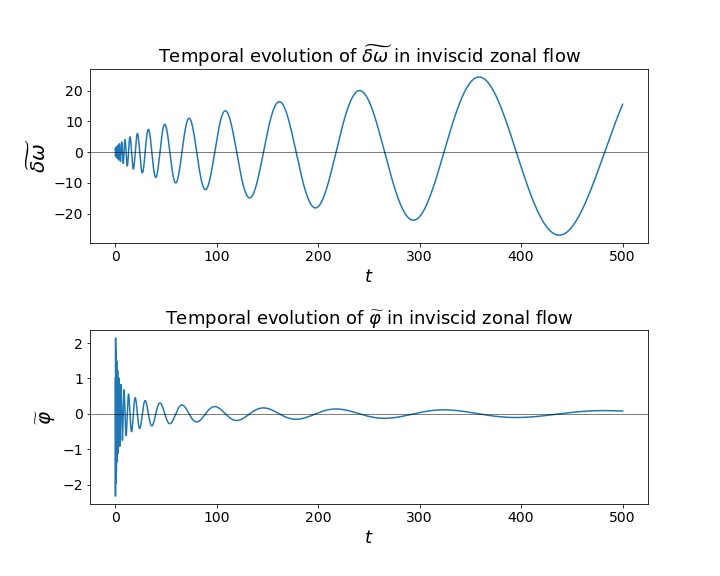}
	\caption{For zonal flow, $\ho$ grows in time, whilst buoyancy force is reduced and $\hphi$ eventually dies out: $k_x(0)=10, k_y(0)=10, N^2=1000, A_z=2$.}
	\label{fig:zfinviscid}
\end{figure}
\begin{figure}
	\centering
	\includegraphics[height=2.5in]{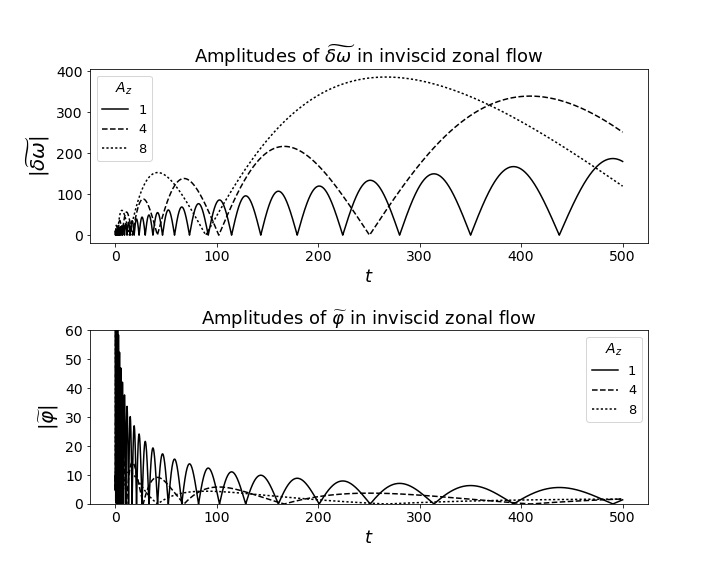}
	\caption{For zonal flow, $\ho$ grows as a power law in time. This growth is more exaggerated as the shearing factor $A_z$ increases; $\hphi$ decays as a power law in time, with the shearing factor amplifying this decay: $k_x(0)=1, k_y(0)=1,  N^2=200$.}
	\label{fig:zfinviscidpow}
\end{figure}
In terms of $\tau = k_x/k_y = A_z t$ and $\alpha = \left| \frac{N}{A_z} \right|$,
Eqs. (\ref{eq9})-(\ref{eq10}) lead to
\begin{equation}
\p_{\tau \tau} \ho + \frac{\alpha^2}{1 + \tau^2} \ho = 0.
\label{eq12}
\end{equation}
In the limit of $\N \gg A_z \gg \nu k_y^2$ ($\alpha \gg 1$), we look for a
WKB solution of the form 
$\ho (\tau) \sim \exp{[\frac{1}{\delta} \int d\tau_1(\psi_0 + \delta \psi_1+  ...) ]}$,
where $\delta \ll 1$ is a small parameter. Plugging this in Eq.~(\ref{eq12}) gives
us $\delta = \alpha ^{-1} \ll 1$, $\psi_0 (\tau_1)= \pm i (1 + \tau_1^2)^{-1/2}$ and $\psi_1(\tau_1) = \frac{\tau_1}{2 (1+ \tau_1^2)}$, and thus
\begin{equation} 
\ho(\tau) \sim  (1+ \tau^2)^{1/4}  e^{\pm i \alpha \theta(t)},
\label{eq14}
\end{equation}
to $O(\alpha^{-1})$.
Here, $\theta(t) =  \sinh^{-1}(\tau)$ and again $\tau = k_x(t)/k_y=A_z { t}$.
By using Eq. (\ref{eq9}), $\vphi= -\frac{ig}{\rho_m} \tn$, and $\hphi = e^{\nu Q} \vphi$ and by assuming the initial condition $\tdomega(t=0)=0$, we find 
\begin{eqnarray}
\tdomega(t) &=& 
-i\frac{g k_y}{\rho_m \N} \left[1 + \tau_0 \right]^{1/4} \left[1 + \tau^2 \right]^{1/4} \sin{(H(t))}
\nonumber \\
&& \times
 e^{-\nu Q_1(t)} \tn(0),
\label{eq16}\\
\tn(t)&=& \frac{ \left[1 + \tau_0^2 \right]^{1/4}}{ \left[1 + \tau^2 \right]^{3/4}}
\left[ \frac{\tau}{2 \alpha} \sin{(H(t))} + \sqrt{1+ \tau^2} 
\cos{(H(t))}\right] 
\nonumber \\
&& \times e^{-\nu Q(t)} \tn(0),
\label{eq17} 
\end{eqnarray}
where $\tau_0=\tau(t=0)=k_x(0)/k_y$, $H(t)=\alpha[\theta(t)-\theta(0)]=\alpha[\sinh^{-1}(\tau(t))-\sinh^{-1}(\tau_0)]$
and $Q_1(t)  = \frac{1}{3A_zk_y}\left(k_x(t)^3 - k_x(0)^3\right) + k_y^2t$.
The $Q_1(t)$ term dominates with leading order $-\frac{1}{3} \nu k_y^2A_z^2t^3 $ for $\nu \ne 0$ as shearing enhances the dissipation over the molecular value. Consequences of this enhanced dissipation are discussed in \cite{KM17}. Furthermore, Eqs. (\ref{eq16})-(\ref{eq17}) show that in the inviscid case, $\ho \propto \tau^{\frac{1}{2}} \propto t^{\frac{1}{2}}$ while $\hphi \propto t^{-\frac{1}{2}}$ for larger $t$. This tendency is seen
in Figs.~\ref{fig:zfinviscid} and \ref{fig:zfinviscidpow} obtained from numerical solutions of Eqs. (\ref{eq2})-(\ref{eq1}). Figs.~\ref{fig:zfinviscid} and \ref{fig:zfinviscidpow} also show that the growth/decay increases with $A_z$, consistent with Eqs. (\ref{eq16})-(\ref{eq17}).  

The density fluctuations decay due to the fact that $\tilde{v}_x  =  \frac{i k_y}{k_x^2 + k_y^2} \tdomega$ decreases with time. For zonal flow, $k_y$ is constant and $k_x^2 + k_y^2$ grows quadratically with time, faster than $\ho^2 \propto t$, so buoyancy diminishes with time. This follows from the total fluctuating energy 
\begin{equation}
E  =  \frac{|\ho|^2}{k_x^2 + k_y^2}  +  \frac{|\hphi|^2}{|N|^2}
 \label{eq18}
\end{equation}
being an adiabatic invariant.

\noindent
{\bf Streamers only}: For the streamer only $\U =(-yA_s,0)$,  Eq. (\ref{eq6}) gives
\begin{equation} 
k_x(t) = k_x(0),\,\,\,\,
k_y(t) = k_y(0) + k_x(0)A_s { t}.
\label{eq19}
\end{equation}
Finding a WKB solution turned out to be tricky in this case. Briefly, we rewrite Eqs. (\ref{eq9}) and (\ref{eq10}) in terms of 
$y=\frac{1}{2} \ln{(1+\tau^2})$ and look for a WKB solution in terms of $y$. The solution which satisfies $\ho(t=0)=0$ can be
found to $O(\alpha^{-1})$ as:
\begin{eqnarray}
\tdomega(t) &=& 
i\frac{g k_x} {\rho_m \N} \left[1 + \tau_0 \right]^{1/4} \left[1 + \tau^2 \right]^{1/4} e^{-\nu Q_2(t)} \tn(0)
\nonumber \\
&& \times \left[- \sin{(R(t))} + \frac{\cos{(R(t))}}{2 \alpha} W(t) \right] ,
\label{eq20}\\
\tn(t)&=& \frac{ \left[1 + \tau_0^2 \right]^{1/4}}{ \left[1 + \tau^2 \right]^{1/4}}
\left[\cos{(R(t))}+\frac{1}{2\alpha \sqrt{1+\tau_0^2}} \sin{(R(t))}\right] 
\nonumber \\
&&\times  e^{-\nu Q_2(t)} \tn(0),
\label{eq21} 
\end{eqnarray}
where $R(t) =\alpha[ \sqrt{1+\tau^2} - \sqrt{1+\tau_0^2}]$, $W(t) = (1+\tau_0^2)^{-\frac{1}{2}} - (1+\tau^2)^{-\frac{1}{2}}$, and 
$
Q_2(t) = \frac{1}{3A_zk_x}\left(k_y(t)^3 - k_y(0)^3\right) + k_x^2 t.
$
\begin{figure}
	\centering
	\includegraphics[height=2.5in]{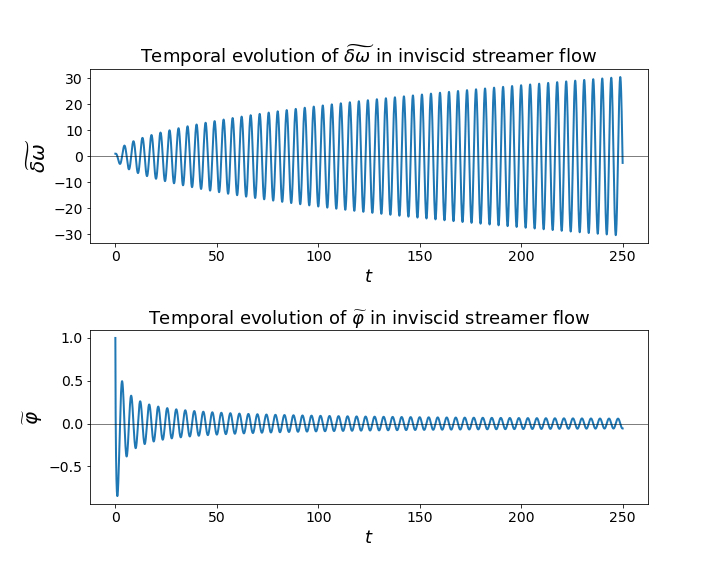}
	\caption{For streamers, $\ho$ grows in magnitude whilst $\hphi$ decays, { similarly to the zonal flow case}. $\omega_f$ approaches a constant value: $k_x(0)=0.1, k_y(0)=0.1, N^2=2,  A_s=30$.}
	\label{fig:stinviscid}
\end{figure}

\begin{figure}
	\centering
	\includegraphics[height=2.5in]{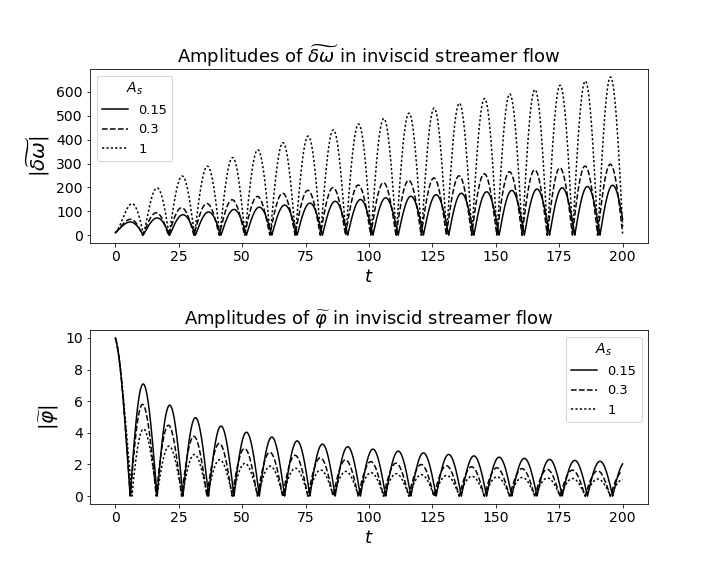}
	\caption{For streamers, $\ho$ and $\hphi$ follow power laws in time, { similarly to the zonal flow case:} $k_x(0)=1, k_y(0)=1, N^2=0.15$.}
	\label{fig:stinviscidpow}
\end{figure} 
In the inviscid case, Eqs. (\ref{eq20})-(\ref{eq21}) show that 
$\ho \propto \tau^{\frac{1}{2}} \propto t^{\frac{1}{2}}$ while $\hphi \propto t^{-\frac{1}{2}}$ for large $t$, similarly to the case of zonal flow only. We confirm this by numerical simulations shown in
Figs. \ref{fig:stinviscid} and \ref{fig:stinviscidpow}.
However, the transport of particles $\langle \rho v_x \rangle$ in zonal flow and streamer is different since
$v_x = \frac{ik_y}{k_x^2 + k_y^2} \propto t^{-2}$ and $t^{-1}$ for large $t$ in zonal flows and streamer cases, respectively.
That is, the transport of particle is less reduced by streamers than zonal flows.

Another marked difference between the streamers and zonal flow case is the frequency $\omega_f$ at which the fluctuations oscillate. For streamers, the frequency remains roughly constant, whilst it always decays in the zonal flow case. This is basically because
$\omega_f  =  |N| \sqrt{\frac{k_y^2}{k_x^2 + k_y^2}}$ when $\nu=D=0$. For zonal flows, $k_x$ grows whilst $k_y$ is constant, leading to
the so-called oscillation death whereby $\omega_f$ decreases towards zero as time progresses. We can see clearly in Fig. \ref{fig:zfinviscidpow}
that the shearing rate $A_z$ is responsible for the decrease in $\omega_f$.  In the streamer case, $k_y$ grows instead whilst $k_x$ remains constant in time. Hence, $\omega_f$ approaches the constant value $\N$, as observed in Figs. \ref{fig:stinviscid} and \ref{fig:stinviscidpow}. This behavior can also be inferred from Eqs. (\ref{eq16}), (\ref{eq17}), (\ref{eq20})
and (\ref{eq21}).

\noindent
{\bf Combined zonal flow and streamers}:
\begin{figure}[t]
	\centering
	\includegraphics[height=2.5in]{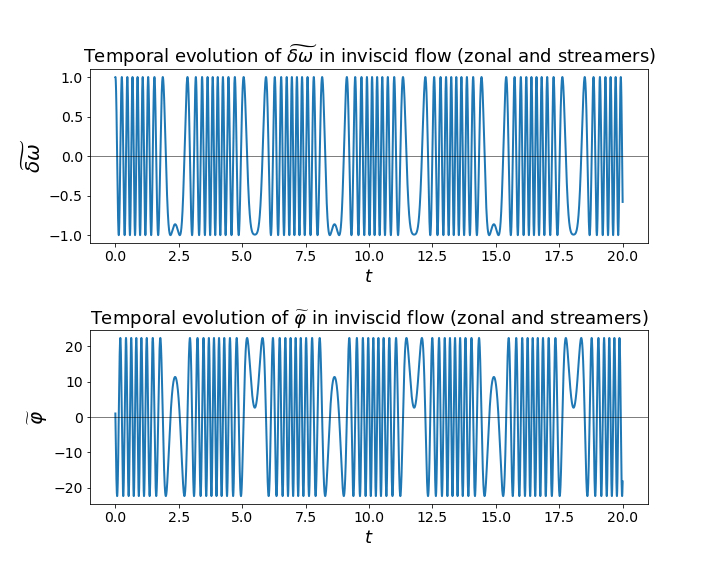}
	\caption{For combined zonal flow and streamers
	with the opposite sign of shear, the amplitudes of both $\ho$ and $\hphi$ do not change: $k_x(0)=1, k_y(0)=1, N^2=1000, A_s=1= -A_z$}
	\label{fig:combflowopp}
\end{figure}
For $\U = (-yA_s, - xA_z)$,
Eq. (\ref{eq6}) gives us $\ddot{k_x}= A_sA_zk_x$, a solution depending on the sign of $A^2=A_sA_z$. To elucidate their effect, we focus on the case where $|A_z|=|A_s|=|A|$. \\
When $A^2<0$, $\kk$ rotates since
\begin{equation}
k_x(t)  =  K \cos(|A|t + \xi),\,\,\,
k_y(t)  =  K \sin(|A|t + \xi),
\label{eq25}
\end{equation}
where $K^2 =  k_x(t)^2 + k_y(t)^2  = k_x(0)^2 + k_y(0)^2$ and $\tan \xi = \frac{k_y(0)}{k_x(0)}$. In this case, we can show that $E$ in Eq. (\ref{eq18}) is now exactly conserved, enabling us to find an exact solution
\begin{equation} 
\tdomega(t) = F  \sin \left[  \left|\frac{N}{A}\right| \sin\left(|A|t+\xi\right) + G \right] e^{-\nu K^2 t},
\label{eq26}
\end{equation}
where
$F  =  \sqrt{ \tdomega(0)^2  +  \frac{K^2}{N^2} {\vphi}(0)^2}$ and 
$G = \sin^{-1} \left(\frac{\tdomega(0)}{F} \right) - \left|\frac{N}{A}\right|\sin{(\xi)}.$
Comparing this with the previous cases, there is no net energy transfer between $\tdomega$ and $\vphi$ on a long time scale;
both fluctuations oscillate with constant amplitudes as demonstrated in Fig.~\ref{fig:combflowopp}.
Furthermore, Eq.~(\ref{eq26}) manifests the oscillation frequency $\omega_f$ at the integer multiples of $A$; $\sin{(\sin{(At)})}$ involves
 the frequency $n A$ for all integer $n$ (see Fig. \ref{fig:combflowopp}).\\
\begin{figure}[t]
	\centering
	\includegraphics[height=2.5in]{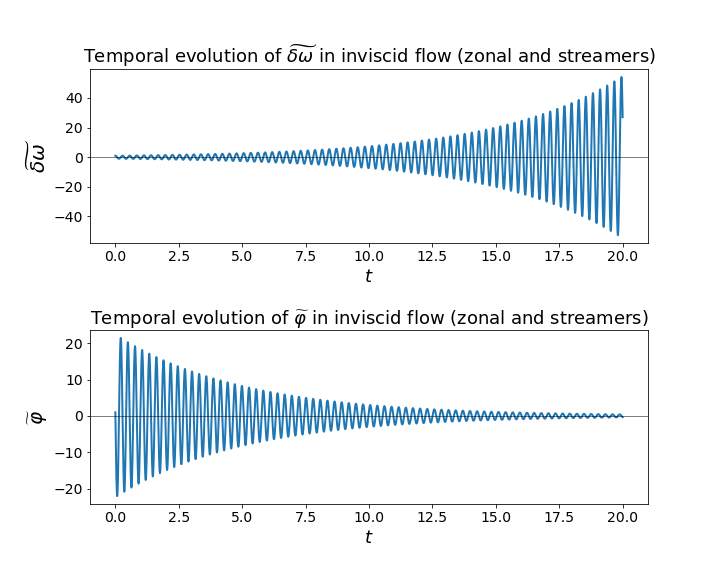}
	\caption{For combined zonal flow and streamers
	with the same sign of shear, $\ho$ and $\hphi$ grow and decay exponentially, respectively: $k_x(0)=1, k_y(0)=1, N^2=1000, A_s=A_z=0.4$.}
	\label{fig:combflowsame}
\end{figure}
In comparison, for $A^2>0$, we have
\begin{equation}
k_x(t)  =  P \cosh(At + \chi),\,\,\,\,
k_y(t)  =  P \sinh(At + \chi),
\label{eq210}
\end{equation}
where $A>0$, $P^2  =  k_x(t)^2 - k_y(t)^2  =  k_x(0)^2 - k_y(0)^2, $ and $\tanh \chi = \frac{k_y(0)}{k_x(0)}$. Eq. (\ref{eq210}) leads to
exponentially increasing and decreasing wave numbers in the two orthogonal directions (see \cite{KM17}). For $At \gg 1$, an inviscid solution is found as $\ho \propto e^{\gamma t}$ where $\gamma = \frac{A}{2} [ 1 + \sqrt{ 1 - \alpha^2} ]$ and
$\alpha=\frac{\N}{A}$. Thus, $\ho$ ($\hphi$) grows (decays) exponentially in time  for all $\alpha$ (see Eq. (\ref{eq18})),
the imaginary part of $\gamma$ giving $\omega_f \to \frac{\N}{2}$ for $\alpha \gg 1$ (consistent with the expectation from $\omega_f = \N \frac{k_y}{\sqrt{k_x^2+k_y^2}}  \to \frac{\N}{2}$ as $k_x \sim k_y \propto e^{At}$). This prediction is confirmed by numerical solutions of Eqs. (\ref{eq2})-(\ref{eq1}) shown in Fig.~\ref{fig:combflowsame}.

In summary, we elucidated the effects of zonal flows and streamers on the evolution of the vorticity and density 
fluctuations in interchange turbulence.
In the inviscid limit, vorticity (density) grows (decays) as a power law $t^{\frac{1}{2}}$ ($t^{-\frac{1}{2}}$) due to streamers or zonal flows.
However, due to the anisotropic stretching of wave numbers, the transport of density is less reduced by streamer than by zonal flow, 
zonal flows leading to oscillation death (reduced oscillation frequency). This highlights different effects of zonal flows and streamers on turbulence regulation. Furthermore, the combined zonal flow and streamer induce oscillations at one frequency with exponentially growing (decaying) amplitude of vorticity (density) or at multiple integer frequencies with constant amplitude, depending on the relative sign of shear. 
Different effects of shear flows, in particular, oscillation death by zonal flows, are likely to persist in forced turbulence, which will addressed in detail in a future extended work.

\end{document}